\documentclass[preprint]{aastex}

\newcommand{\kms}{km\,s$^{-1}$}

\newcommand{\kmsMp}{km\,s$^{-1}$\,Mpc$^{-1}$}
\newcommand{\mJybeam}{mJy\,beam$^{-1}$}
\newcommand{\atms}{atoms\,cm$^{-2}$}
\newcommand{\msun}{{$M_\odot$}}
\newcommand{\lsun}{{$L_\odot$}}
\newcommand{\hi}{{\ion{H}{1}}}
\newcommand{\HI}{{\ion{H}{1}}}
\newcommand{\degree}{{$^\circ$}}

\newcommand{\HII}{{\ion{H}{2}}}
\newcommand{\NII}{{\ion{N}{2}}}

\input psfig

\slugcomment{Submitted to AJ, \today}

\shorttitle{\ion{H}{1} in low-luminosity
E/S0 and S0 galaxies}
\shortauthors{Sadler et al.}

\begin{document}

\title{\ion{H}{1} in four star--forming low-luminosity 
E/S0 and S0 galaxies\footnote{Based on observations with the Australia Telescope
Compact Array (ATCA), which is operated by the CSIRO Australia Telescope
National Facility}}

\author{Elaine M. Sadler}
\affil{School of Physics, University of Sydney, NSW\,2006, Australia}
\email{ems@physics.usyd.edu.au}

\author{Thomas A. Oosterloo\altaffilmark{2,3}}
\affil{Istituto di Fisica Cosmica, CNR, Via Bassini 15, 20133 Milan, Italy}
\email{oosterloo@nfra.nl}

\author{Raffaella Morganti\altaffilmark{2}}
\affil{Istituto di Radioastronomia, CNR, via Gobetti 101, 40129 Bologna,
Italy}
\email{rmorgant@ira.bo.cnr.it}

\and

\author{Amanda Karakas}
\affil{Australia Telescope National Facility, CSIRO, PO Box 76, 
Epping, NSW 2121, Australia}
\email{}

\altaffiltext{2}{Australia Telescope National Facility, CSIRO, PO Box 76, 
Epping, NSW 2121, Australia}
\altaffiltext{3}{Netherlands Foundation for Research in Astronomy, Postbus 2,
7990 AA,  Dwingeloo, The Netherlands}

\begin{abstract}
We present \hi\ data cubes of four low--luminosity early--type (E/S0 and
S0) galaxies which are currently forming stars.  These galaxies have absolute
magnitudes in the range $M_{\rm B} = -17.9$ to $-19.9$ ($H_\circ$ = 50
\kmsMp).  Their \hi\ masses range between a few times $10^8$ and a few
times $10^9$ $M_\odot$ and the corresponding values for $M_{\rm HI}/L_{\rm B}$
are between 0.07 and 0.42,  so  these systems are \HI\ rich for their
morphological type.  In all four galaxies, the \HI\ is strongly centrally
concentrated with high central \HI\ surface densities, in contrast to what is
typically observed in more luminous early--type galaxies.
Star formation is occurring only in the central regions.\\  
In two galaxies (NGC\,802 and ESO\,118--G34), the kinematics of the \HI\
suggests that the gas is in a strongly warped disk, which we take as evidence 
for recent accretion of \HI.  In the other two galaxies (NGC\,2328 and 
ESO\,027--G21) the \HI\ must have been part of the systems for a considerable
time. The \HI\ properties of low--luminosity early--type galaxies appear to be
systematically different from those of many more luminous early--type galaxies, 
and we suggest that these differences are due to a different evolution of the
two classes. 

The star formation history of these galaxies remains unclear.  Their $UBV$ 
colours and H$\alpha$ emission--line strengths are consistent with having 
formed stars at a slowly--declining rate for most of the past $10^{10}$ years.  
If so, their star formation history would be intermediate between late--type
spiral disks and giant ellipticals.  However, the current data do not rule out
a small burst of recent star formation overlaid on an older stellar population.

Three of the galaxies have weak radio continuum emission, and the ratio of the
far--infrared (FIR) to radio--continuum emission is very similar to that of
spirals of similar FIR- or radio luminosity.  We find that, except in the
largest galaxy observed, the radio--continuum emission can be accounted for
solely by thermal (free--free) emission from \HII\ regions, with no 
non--thermal (synchrotron) disk component.  Thus, although these galaxies have
gaseous disks, a disk magnetic field may be very weak or absent. 
\end{abstract}

\keywords{galaxies: kinematics and dynamics -- radio emission lines}

\section {Introduction}

Disentangling the star--formation history of early--type galaxies is not
straightforward.  Such galaxies have traditionally been regarded as old,
gas--poor stellar systems, yet we now know that they have a complex,
multi--phase interstellar medium (e.g.\ Knapp 1999) and often contain
detectable (and occasionally large) amounts of neutral hydrogen (Knapp et al.\
1985).  The standard view is that such galaxies are dominated by an old
stellar population which probably formed quite rapidly.  This appears to be
true for the most luminous galaxies, but not necessarily for those of lower
luminosity for which the star formation histories could be different (e.g.\
Faber et al.\ 1995; Worthey 1996). Differences in star--formation histories
are indicated by, for example, different abundance ratios as function of 
luminosity
(e.g.\ Worthey et al.\ 1992, Matteucci 1994).  Such differences may point to
longer periods of star formation in smaller early--type galaxies, with
corresponding differences in the way the ISM is enriched.  Such more extended
star formation could be related to the disks that become more important
towards lower luminosities (e.g. de Jong \& Davies 1997).  Many
low-luminosity early--type galaxies show direct evidence for ongoing star
formation in their central regions in the form of \HII\--region emission--line
spectra (Phillips et al.\ 1986).

Differences in the star--formation history of large and small early-type
galaxies are mirrored by other differences between the two groups.  In
general, low--luminosity galaxies are less well studied than their giant
counterparts (probably because they are under-represented in magnitude--limited
galaxy samples); but several global properties of early--type galaxies appear
to change at an absolute magnitude of $M_{\rm B} \sim -19$ to $-20$.  Luminous
ellipticals often have boxy isophotes and anisotropic velocity dispersions,
while smaller ellipticals tend to be rotationally flattened (e.g., Kormendy \&
Bender 1996) and at large radii they contain disks very similar to S0 galaxies
(Rix, Carollo \& Freeman, 1999). The two groups also have different radio and
X--ray properties, with luminous galaxies being more likely to have X--ray
coronae (Canizares et al.\ 1987) and active nuclei (Sadler et al.\ 1989).  In
the last few years, HST observations have shown that a relation exists between
the central structure and global galaxy properties like rotation (e.g.\ Lauer
et al.\ 1995).

Many of these differences can be understood in terms of the different amounts
of gas present during a galaxy's formation and evolution.  In many models for
galaxy formation (e.g.\ Kauffmann 1998, Baugh et al.\ 1998), the gas supply is
a key factor in determining the characteristics of early--type galaxies.  It
is therefore important to know the properties of the gas in elliptical and S0
galaxies over a wide range of luminosities, and to study how these properties
relate to other characteristics of these galaxies.

To study in more detail the relation between gas content and other structural
parameters, we are undertaking a program to study the \HI\ properties in a
large number of early--type galaxies (Morganti et al.\ 1997a, 1997b, Oosterloo
et al.\ 1999a, 1999b, Morganti et al.\ 1999).  In this paper, we present \HI\
observations of four gas--rich, low--luminosity early--type galaxies that were
listed by Phillips et al.\ (1986) as having \HII\ region--like optical spectra.

Throughout this paper, we use $H_\circ$ = 50 \kmsMp\ unless otherwise stated.

\section{Four low--luminosity early--type galaxies with \HII\--region spectra} 

Phillips et al.\ (1986) listed seven low--luminosity E/S0 and S0 galaxies with
optical spectra characteristic of \HII\ regions.  We have used the Australia
Telescope Compact Array (ATCA) to image four of these galaxies (NGC\,802, 
NGC\,2328, ESO\,118--G34 and ESO\,027--G21) in the \HI\ line.  A fifth galaxy, 
NGC\,1705, has already been observed in \HI\ by Meurer et al.\ (1992, 1998).  
We briefly describe the four galaxies we observed, and summarize their 
properties in Table 1. 

\noindent {\sl NGC\,802}. 
This is an isolated E/S0 galaxy of absolute magnitude $M_B = -18.0$
which has been very little studied.  A spiral
galaxy to the north--east, ESO 052--G14, is at higher redshift and not
associated (Sadler \& Sharp 1984).  NGC 802 is listed as a UV--bright galaxy
by Coziol et al.\ (1997).

\noindent {\sl ESO\,118--G34}.
This galaxy is classified as an S0 galaxy that lies in a small group of
galaxies which includes the spirals NGC 1672 and NGC 1688 (Garcia 1993).  
Its absolute magnitude is $M_B = -17.9$.  It has been imaged in 
H$\alpha/$[\NII] by Buson et al.\ (1993).  The galaxy appears to be actively
forming stars, with several bright \HII\--region complexes in the central
$\sim$20 arcsec. 

\noindent {\sl NGC\,2328}.
This is an isolated E/S0 galaxy of $M_B = -18.7$.  It has been detected in CO
by Lees et al.\ (1991) and Wiklind et al.\ (1995).  A narrow--band image in
H$\alpha/$[\NII] (Sadler 1987) shows extended emission--line gas in the central
9\,arcsec, distributed in a clumpy, ring--like structure.

\noindent {\sl ESO\,027--G21}.
This is an S0 galaxy ($M_B = -19.9$) which lies in a small group of galaxies,
and has not previously been studied in detail.  Optical images suggest that
the galaxy may have a weak spiral arm extending to the south--east.

\section {Radio observations}

\subsection{Single--dish \hi\ observations at Parkes} 

Three of our four galaxies were observed in 1988 with the Parkes telescope 
by Sadler and J.B.\ Whiteoak. 
The Parkes 64-m telescope has a beam of 14.7\,arcmin at 1.415\,GHz.  The 1988
observations used two cryogenically cooled FET receivers connected to
horizontal and vertical polarized ports of the feed, with noise temperatures
of $\sim$60\,K.  NGC~2328 was observed with a bandwidth of 5\,MHz; and NGC~802
and ESO~027--G21 with a 10~MHz bandwidth.  The 1024-channel autocorrelator was
used in $2\times 512$ channel mode, giving a velocity scale of 2.1 \kms\ per
channel for NGC~2328 and 4.2 \kms\ per channel for the other two galaxies.  
The final resolution (after smoothing) is roughly 4 and 8 \kms\ respectively. 

Table 2 summarizes the results and the derived \hi\ masses, while Figure 1
shows the Parkes single--dish profiles. 

\subsection {ATCA Observations }

We used several different configurations of the Australia Telescope Compact
Array (ATCA) to image the four galaxies in \hi.  Table 3 summarizes the
observations.  Except for one observation (NGC~802 with the 375-m array), 
we used a narrow (8 MHz) band with 512 velocity channels.  The band was
centered on the optical velocity of each galaxy, and the final velocity
resolution (obtained after hanning smoothing the data) is $\sim$6 \kms.
We generally chose compact configurations of the ATCA (375-m and 750-m arrays) 
to allow us to investigate the full extent of the \hi\ emission.  As a result,
the spatial resolution of our \hi\ images is typically $\sim$50 arcsec. 
Two of the galaxies were also observed with 1.5-km arrays to improve the 
spatial resolution in the centre. 

We used the second IF for simultaneous continuum observations with a bandwidth
of 128 MHz and 33 channels.  The observed continuum frequencies are given in
parentheses in Table~3.  For ESO~118--G34, two continuum frequencies (13 and 20
cm) were measured. 

We observed each galaxy for about 12 hours in each run, with calibrators
observed every hour to monitor the gain changes.  The flux density scale was
set by observations of PKS 1934--638, for which we adopted a flux density of
14.9 Jy at 1400 MHz.  This source was also used as the bandpass calibrator. 

\subsubsection{Neutral Hydrogen Data}

The spectral data were calibrated with the MIRIAD package (Sault et al.\
1995), which has several features particularly suited for ATCA data.  The
continuum subtraction was also done in MIRIAD, by using a linear fit through
the line--free channels of each visibility record and subtracting this fit
from all the frequency channels (i.e.\ UVLIN).  We already knew that none of
the four galaxies contained a strong continuum source (Sadler 1984; see also
below), so continuum subtraction did not cause any complications for the line
data.  An interference spike generated by the ATCA data acquisition system is
present at 1408~MHz, but since this frequency is outside the velocity range of
interest, it does not affect our data.

For NGC~802 and ESO~118--G34, which were observed with more than one ATCA
configuration, we combined the data from different runs after calibration and
continuum subtraction.  The final reduction (i.e.\ CLEAN and moment analysis
of the data cube) was done using GIPSY (van der Hulst et al., 1992).  The
final cubes were made with both natural and uniform (or robust) weight.
Table~4 lists the corresponding rms noise and size of the restoring beams.

As expected from the Parkes observations, we detect \hi\ in all four galaxies 
and Table 2 gives the \hi\ fluxes.  
The total intensity and intensity--weighted mean--velocity fields of the \hi\
emission were derived from a data cube by smoothing the original data cube
spatially to a resolution about twice as low as the original.  The smoothed
cube was used as mask for the original cube: pixels with signal below
$3\sigma$ in the smoothed cube were set to zero in the original cube (van
Gorkom \& Ekers 1989).

The total \hi\ fluxes measured at the ATCA are typically 10--30\% lower than
those measured at Parkes, which may imply that some very extended low--level 
\hi\ has been resolved out with the ATCA or removed by the 3$\sigma$ clipping. 
It is also possible that the single--dish fluxes could be slightly too high
because of miscalibration in the 1988 data.  Data from the Parkes HI multibeam
survey (Barnes et al., in preparation), due for release in 2000, will provide
an independent check of the single--dish data. 

\subsubsection{Continuum Data}

The continuum data from the ATCA were also reduced using MIRIAD, and Table~5
summarizes the results.  Data from the second IF were used (see Table~3),  
except for the 1.5-km array observation of ESO\,118--G34, where the second 
IF was set to 2.4\,GHz and a 1.3\,GHz flux density was measured from 
emission--free channels in the line data (first IF). 

For each galaxy in Table 5, we give two continuum measurements.  The first
uses all the available data, except baselines to the 6-km ATCA antenna.  This
gives a typical beam of 20--40 arcsec diameter, and hence a good estimate of
the total flux density since all the sources are likely to be unresolved at
this resolution.  To get the highest possible spatial resolution (about 8
arcsec), we made a second set of images by including the baselines to the 6-km
ATCA antenna as well. This gives a very patchy $uv$ coverage because of the
big gap between the long and short baselines, and the emission may not be
imaged very well, but a comparison of the fluxes derived from the low- and
high-resolution images gives us some information on the likely spatial extent
of the continuum emission.

Three of the four galaxies were detected as continuum sources, but the
continuum emission was weak and either unresolved (ESO\,027--G21) or barely
resolved (NGC\,2328 and ESO\,118--G34) in the higher--resolution images.  In
all three detected galaxies, the radio continuum source is centered on the
galaxy's nucleus and almost all the continuum emission arises from within the
central 5--10\,arcsec.

For ESO~118--G34, we have continuum data at both 1.3 and 2.4\,GHz, and measure 
a spectral index of $-0.6$.  NGC~2328 was detected at 5\,GHz with a flux density
of 2.5\,mJy by Sadler et al.\ (1989) using the VLA B/C array.  This implies a 
steep spectral index of $-0.9$, but the true spectral index may be flatter than
this since it is possible that some extended 5\,GHz emission was resolved out
by the VLA.  For the remaining galaxies, the only other continuum measurements
available are upper limits at 11\,cm (Sadler 1984).

\section {Results}

Figures 2 and 3 show the \HI\ distribution and velocity field, obtained using
natural weighting, for each of the four galaxies observed.  Figure 4 shows the
\HI\ total--intensity distributions and velocity fields obtained by using
uniform (or robust) weighting, with slightly higher spatial resolution (see
Table 4 for the beam sizes).  The low resolution total \HI\ images can give an
indication of the overall extent of the \HI\ emission compared to the optical
object, while the higher resolution data gives more information about the
kinematics of the \HI. Table 6 summarizes the main \HI\ properties derived 
from the ATCA images. 

In Figure 5 we show the position--velocity plots along the kinematical major
axis of the \HI\ of the four galaxies.  For NGC~802 the plot obtained from the
high resolution data cube is also shown.  In the following section, we briefly
summarize the characteristics of the \hi\ emission for each object.

\subsection{Notes on individual galaxies }

\subsubsection{NGC 802}

The most striking characteristic of the \HI\ in NGC 802 is its kinematics: 
the velocity field shows that the rotation axis of the bulk of the \HI\ appears
to be quite mis-aligned with the optical minor axis. However, the
position--velocity diagrams of this galaxy show that the \HI\ in this galaxy 
may not be in a single, flat disk--like structure and that the velocity field,
given the broad \HI\ profiles and the relatively low spatial resolution, does
not give a complete description of the kinematics of the \HI.  The
high--resolution version of the position--velocity diagram, taken along the
kinematical major axis, shows that the brightest \HI\ rotates roughly about 
the optical major axis, but there is also fainter \HI\ at higher projected
velocities which appears to lie closer to the center of the galaxy (the ridge
of \HI\ seems to curve back to the center at velocities away from systemic). 
This is suggestive of a warped \HI\ structure like that observed, for example, 
in the luminous dust--lane elliptical NGC\,5266 (Morganti et al.\ 1997a).  
The ionized gas also shows very
little rotation along the optical major axis. Figure 6 shows the 
ionized gas velocities measured from an optical spectrum (see \S 5.1) taken
along PA $145^\circ$, roughly aligned with the optical major axis.  A very
shallow velocity gradient is observed.  It is possible that NGC 802 is forming
a minor--axis dust lane or a polar ring.  NGC 802 shares this characteristic
with the small elliptical galaxy NGC 855 (Walsh et al.\ 1990) and possibly
also with ESO118--G34 (see below).  This strongly suggests that a significant
fraction of the \HI\ in NGC~802 has been accreted relatively recently.  The
kinematic timescale ($V/R$) in NGC 802 is of the order of a few times $10^8$
years. Given that the \HI\ distribution is not very chaotic, the accretion
event probably occurred of the order of $\sim$$5 \times 10^8$ years ago.

The radial surface--density profile of the \HI\ in NGC 802 (Figure 7)  is
quite peaked towards the center, with the central density being about $4.5$
\msun ${\rm pc}^{-2}$. Given the limited spatial resolution of the \HI\ data,
the true central surface density is likely to be even higher. This value is
much higher than typically found in more luminous early--type galaxies (that
usually have values in the range 0--2 \msun ${\rm pc}^{-2}$ e.g.\ Oosterloo
et al.\ 1999a, 1999b), and is consistent with the fact that star formation is
occurring in the center of NGC 802.

\subsubsection{ESO~118--G34}

The position--velocity diagram of the \HI\ in ESO~118--G34, taken along PA
160\degree, is qualitatively similar to that of the \HI\ in NGC 802. Also
here, very broad profiles are observed and the overall shape of the \HI\ in
this diagram is also S-shaped. Given the round shape of the optical isophotes,
it is not possible to determine the orientation of the rotation axis relative
to that of the optical galaxy. Although there is an overall rotation pattern
in the \HI, the velocity field shows deviations from that what expected for a
flat disk, and also the \HI\ distribution also has some irregular features. 
In the inner parts, the kinematical major axis seems to be along PA 160\degree,
while at larger radii it seems to be more like 45\degree.  It appears that also
in ESO~118--G34 the \HI\ is likely to be in a strongly  warped structure.  
The range of velocities observed suggests that we are viewing the \HI\ from
close to face--on. Hence, small changes in the orientation of the \HI\ disk
lead to relatively large signatures in the velocity field.  The \HI\
distribution in ESO 118--G34 is quite extended, but strongly peaks in the
center, with central surface densities very similar to those observed in NGC
802.

\subsubsection{NGC~2328}

Of the four galaxies we observed, NGC 2328 has perhaps the most regular \HI\
distribution. The \HI\ is in a regularly rotating disk, aligned with the
optical image of the galaxy. As in the other galaxies, the \HI\ distribution
peaks quite markedly in the center, although it seems that the central \HI\
density could be  somewhat lower than in NGC 802 and ESO 118--G34.

Although the spatial resolution is quite low, the position--velocity diagram
suggests that the rotation velocity does not decrease in the outer part of the
\HI\ disk. This would indicate the presence of a dark matter halo.  We can
make a very rough estimate of the mass--to--light ratio in this galaxy.  
The projected rotation velocity at 1.5\,arcmin radius is about 50 \kms.  
Assuming circular orbits and a spherical mass distribution, this implies 
$M/L_B \sim 1.1/\sin^2 i$ at this radius. Although we are limited by the
resolution of our data, we estimate that the \HI\ disk has an inclination of
about 30\degree. The optical axis ratio gives a lower limit to the inclination
of about 25\degree, so the \HI\ disk in NGC 2328 is observed relatively
face--on and as a result the mass estimate is quite uncertain.  
If we adopt 30\degree\ for the inclination of the \HI\ disk, we find 
$M/L_B \sim 4.6$ at about $5R_{\rm e}$ (i.e.\ 6~kpc from the center).  
Higher resolution observations, complemented with optical data, would be needed
to make a proper study of the mass distribution in this galaxy. 

\subsubsection{ESO~027--G21}

In ESO 027--G21, as in NGC 2328, the \HI\ appears to be in a regularly
rotating disk aligned with the optical major axis of the galaxy.  The 
velocity field shows that the kinematical position angle changes with radius,
indicative of a warped \HI\ disk.  Also the faint extensions of the \HI\
distribution to higher projected velocities at large radii in the
position--velocity map shows this.

Also in this galaxy, the surface density profile peaks at the center with a
relatively high surface density of almost 6 \msun ${\rm pc}^{-2}$. The
value derived from the natural weighted cube is basically the same as that
derived from the cube made with robust weighting, indicating that this value
is probably not much affected by the low spatial resolution.  An optical
spectrum taken along the major axis (Figure 6, see also \S 5.1) shows that the
rotation velocities rise quite steeply in the center, and that the level of
rotation of the inner \HI\ disk of 65 \kms\ is already reached at a radius of
a few arcseconds (Figure 6).  The position--velocity map (Figure 5) shows,
contrary to the other three galaxies observed, a minimum in the center.  Given
that the radial \HI\ density profile (in space) peaks in the center, this
implies that the rotation curve in this galaxy rises more quickly than in the
other galaxies, as is observed in the optical spectrum. 

As for NGC 2328, the \HI\ data for this galaxy allow us to estimate the
mass--to--light ratio.  With the same assumptions, taking a projected rotation
velocity of the outermost \HI\ at 1.5\,arcmin radius of about 100 \kms, we find
$M/L_B \sim 3.3/\sin^2 i$ at a radius of about $5 R_{\rm e}$ in ESO~027--G21.
Taking the optical axis ratio to make an estimate of the inclination, we find
$i = 40$\degree and hence $M/L_B \sim 8$.

\subsection {Distribution and Kinematics of the \HI\  } 

The kinematics and distribution of the \HI\ in the four galaxies can be
summarized as follows:

\begin{itemize}

\item 
All four galaxies have an extended \HI\ distribution, typically extending 
to $5R_{\rm e}$

\item 
The \HI\ surface density is strongly peaked at the center, with central 
surface densities of at least 4 \msun ${\rm pc}^{-2}$. These high
surface densities are related to the central star formation regions 
observed in all four galaxies. 

\item 
The \HI\ lies in a rotating, disk--like structure, although the exact \HI\
structure varies from galaxy to galaxy. In two of the galaxies (NGC 802 and
ESO 118--G34) the \HI\ is likely to be strongly warped, and in NGC 802 the
kinematical major axis of the \HI\ is even perpendicular to the optical major
axis. This suggests that in these two galaxies a significant fraction of the
\HI\ has been accreted relatively recently (of the order of $5 \times 10^8$ yr
ago). In the two other galaxies (NGC 2328 and ESO 027--G21) the \HI\ appears
to have a more regular distribution and kinematics, although in ESO
027--G21 the outer \HI\ disk also shows a  mild warp. 

\end{itemize}

Most of these results are in good agreement with the findings of Lake et al.\
(1987) for the four low--luminosity ellipticals they studied.  Lake et al.\
also find the \HI\ distributed in relatively regularly rotating disks
extending well beyond the optical galaxy, with some indications that at
least some of the \HI\ has been accreted relatively recently, or is in fact
accreting. They also find that the \HI\ in low--luminosity early--type galaxies
is more centrally concentrated than in more luminous early--type galaxies.
 
In two galaxies (NGC 2328 and ESO 027--G21) we can make a very rough estimate
of the mass--to--light ratio, and we find values of 5 -- 8 \msun/\lsun in the
$B$ band, at $\sim$5 $R_{\rm e}$. Although these values are quite uncertain,
it does appear that they are significantly lower (by a factor 2--4) than what
has been found for more luminous early--type galaxies (e.g.\ Morganti et al.\
1999), where typical values of 20 \msun/\lsun\ are found at
$\sim$5 $R_{\rm e}$. In general, lower values for $M/L$ are found for
low--luminosity early--type galaxies (see e.g.\ J\o rgensen, Franx and Kj\ae
rgaard 1996). Given the star formation we observe in the four galaxies, this
difference is likely to be due to the different stellar populations in the two
groups of galaxies rather than a difference in dark matter content.  In
fact, in the $V$ band the differences in $M/L$ are smaller (by roughly a
factor 1.6), given that our low--luminosity galaxies have bluer colours
than the more luminous galaxies (assuming that ESO 027--G21 has similar
colours to the other three galaxies). Hence the low values for $M/L_{\rm B}$ 
probably arise from the very blue colors of these galaxies, as can be 
seen from Table 1. 

In these two galaxies, the fastest rotating \HI\ is observed at the largest
radii, which would indicate the presence of dark haloes, but higher resolution
data are needed to determine to what extent this is due to warping of the
outer \HI\ disk, as seems to be the case in ESO 027--G21. The rotation curve
of ESO 027--G21 appears to be flat from a radius of a few arcseconds to about
1 arcminute where the disk appears to become more edge on. Given that the
effective radius of this galaxy is 19 arcseconds, this is suggestive for the
presence of dark matter.

\subsection{\ion{H}{1} emission from other galaxies in the field}

In two cases we detected \HI\ from another galaxy in the fields of our program
galaxies.  Figure 8 shows \HI\ total--intensity images for these two objects. 

In the NGC~802 field, we detected \HI\ emission from AM~0155--675 (J2000.0
position $\alpha = 01^{\rm h} 57^{\rm m} 04^{\rm s}$, $\delta = -67^\circ
42^\prime 54^{\prime\prime}$), which lies $\sim 15$ arcmin NW of NGC~802. This
object is classified as a pair of galaxies by Arp\& Madore (1987).  We measure
an \hi\ systemic velocity of 1385 \kms $\pm 10$ \kms, though it is not clear
whether the \hi\ is associated with both members of the pair, or only one.  
No optical velocity is catalogued. 

An anonymous galaxy is also detected  5 arcmin south of ESO~027--G21. The \HI\
systemic velocity of this galaxy is 2455 \kms $\pm 10$ \kms, i.e.\ very close
to that of ESO 027--G21.  Corwin et al.\ (1985) suggest that this galaxy may be
interacting with ESO 027--G21, but the \HI\ data do not show any indications
for such an interaction.

\section{The interstellar medium and star formation in the four galaxies }

Table 6 summarizes the known properties of the interstellar medium (ISM) in the
four galaxies.  In general, early--type galaxies have a complex, multi--phase
ISM (e.g.\ Knapp 1999) which can include cold (CO, H$_2$), cool (\hi), warm
(HII) and hot (X--ray) components. 

Although there is no information about the X--ray properties of the four
galaxies studied here, such low--luminosity galaxies would not be expected to
contain detectable amounts of hot X--ray gas (e.g.\ Fabbiano 1989) and it seems
reasonable to assume that cooler gas dominates the ISM. 
  
There is also very little known about the molecular gas content 
of these galaxies.  CO has been detected in NGC~2328 by Lees et al.\ (1991) 
and Wiklind et al.\ (1995).  The observations by Lees et al.\ used the 
10.4\,m telescope of the Caltech Submillimeter Observatory on Mauna Kea 
to observe the $^{12}$CO(2--1) line at 230.5\,GHz.  The telescope half--power 
beamwidth at this frequency was 32\,arcsec, and they derived a molecular 
hydrogen mass of $6.6(\pm1.0)\times10^7$\msun\ for NGC\,2328 (converted 
to $H_\circ$ = 50\kmsMp), based on the 
generally--adopted conversion relation between CO and H$_2$. Wiklind et al.\ 
used the 15\,m SEST telescope in Chile to observe NGC\,2328 in the J=1--0 
CO line at 115\,GHz, where the SEST half--power beam was 44\,arcsec. 
They derived an H$_2$ mass of $2.9\times10^8$\msun\ for NGC\,2328, i.e. about 
four times larger than that measured by Lees et al. Since their measurements 
used a slightly larger beam this may mean that the CO is spatially 
extended over at least the central 2--3\,kpc, or may simply reflect 
uncertainties introduced by the assumption of a H$_2$/CO conversion ratio. 
The \hi\ mass in the central 30--40\,arcsec of NGC\,2328, based on the 
radial profile shown in Figure 7, is around 3$\times10^7$\msun, so that 
molecular hydrogen appears to dominate the ISM of NGC\,2328 in the 
central 2\,kpc where star formation is taking place.  

Because the CO measurements are made with single dishes, little is known 
about the kinematics of the molecular gas.  Lees et al.\ measure FWHM 
= 112\,\kms from their CO profile, i.e.\ almost identical to the FWHM of 
111\,\kms\ measured from the Parkes \hi\ profile in Figure 1.  Wiklind et 
al.\ measure a much larger FWHM of 280\,\kms\ from the SEST data, and the 
reason for this is unclear. 

In summary, it is clear that molecular gas is an important constituent of 
the ISM in these galaxies, and further CO data would be very valuable.  

In the four galaxies in Table 6, the \hi\ mass to blue luminosity ratio
($M_{\rm HI}/L_B$) has values around 0.1 to 0.4, which are more typical for
spiral galaxies than for early--type galaxies (Bregman et al.\ 1992).

All four galaxies were detected by IRAS at 60 and 100\,$\mu$m (Knapp et al.\
1989) and we can use the far--infrared flux densities to estimate the mass of
dust which is present (Roberts et al.\ 1991).  The dust masses inferred in
this way are quite low (typically a few times $10^5$\msun), and the ratios
$M_{\rm HI}/M_{\rm dust}$ range from 1500 to over 6000. A ratio of about
1000 is often observed in early--type galaxies (Henkel \& Wiklind,
1997), although some galaxies do have much higher ratios. These high ratios
might mean that the abundances in the ISM are low because much of the
\hi\ is accreted recently and is primordial. It is also quite possible that 
much more dust is present but that it is too cool to emit in the IRAS bands
(e.g.\ Young et al.\ 1989).

\subsection{Location of the star--forming regions }

Two of the galaxies studied (ESO 118--G34 and NGC 2328) show distinct, 
clumpy HII regions in CCD frames (Buson et al.\ 1993; Sadler 1987).  
In both these galaxies, the H$\alpha$ emission is confined to the central
1--2\,kpc. 

In ESO\,118--G34, narrow--band CCD images show that the line--emitting region
has a diameter of about 1.3\,kpc, or 17 arcsec (Buson et al.\ 1993), and the
mass of ionized gas is estimated as 5.1$\times10^4$\,\msun.  In NGC\,2328
the emitting region is somewhat smaller, with a diameter of 0.6\,kpc or 8
arcsec (Sadler 1987), but the mass of ionized gas is similar at
6.8$\times10^4$\,\msun. In both cases the star--forming region is very small 
compared to the \HI\ disk.

For the other two galaxies (NGC 802 and ESO 027--G21), no narrow--band CCD
images are available.  As a result, little is known about the structure of the
star--forming regions, except that they are centrally concentrated (because the
Phillips et al.\ (1986) spectra show emission lines from the nucleus).  This
is confirmed by optical spectra taken by us in July 1997 at the ANU 2.3-m
telescope at Siding Spring, and shown in Figures 6a and 6b.  For each of these 
two galaxies, a single spectrum was taken with the slit aligned along the
optical major axis.  In both galaxies, strong emission lines were detected,
with emission--line ratios typical of star--forming regions. This is in 
contrast with  more luminous galaxies, where, if ionized gas is detected, the
spectrum is usually LINER--like and the emission is not associated with star
formation (e.g.\ Phillips et al.\ 1986, Goudfrooij 1999). 
In NGC 802 the gas extends to about 6 arcsec from the nucleus
(though, as noted earlier, the misalignment between the kinematic axes of
stars and gas in this galaxy means that the spectrograph slit was probably not
aligned with along the major axis of the gas disk).  In ESO 027--G21, the gas
extends somewhat further from the nucleus (10--12 arcsec) but is still
centrally concentrated.

Hence, we confirm that star formation is occurring in all galaxies and that 
the brightest star--forming regions are confined to the innermost regions of
the central stellar bulge. The fact that the star formation is occurring in 
the center is consistent with the strong central peaks we observe in the \HI\
distributions.

\subsection{Star formation rates }

We can use the H$\alpha$ emission--line fluxes observed by Phillips et al.\
(1986) to estimate the current star formation rate in each of our four
galaxies. Although the Phillips et al.\ measurements are from slit spectra, 
the H$\alpha$ fluxes derived in this way give a reasonable estimate of the
total value because the emission region is so centrally concentrated. Buson 
et al.\ (1993) discuss this further, and compare slit and CCD total fluxes for
several galaxies. 

We use the relations between H$\alpha$ luminosity and star formation rate 
given by Kennicutt (1983) to derive the star formation rate in solar masses 
per year for (a) stars more massive than 10\,M$_\odot$ and (b) all stars. 
Kennicutt assumes an initial mass function of the form: 

\begin{center} 
\noindent
$\phi(m) \propto m^{-1.4}$  (0.1 $\leq m \leq$ 1\,M$_\odot$), and \\
$\phi(m) \propto m^{-2.5}$  (1 $\leq m \leq$ 100\,M$_\odot$). 
\end{center} 

Table 7 shows the results.  The derived star formation rates are relatively
low in absolute terms. We find values lower than 0.5\,\msun\ per year, 
compared to $\sim$4\,\msun\ per year for spirals like our own Galaxy and up to
50\,\msun\ per year for extreme starburst galaxies. These are clearly not 
`starburst' galaxies, but relatively quiescent systems.

However, the galaxies in Table 7 are only about one--tenth as luminous as the
bright spiral galaxies studied by Kennicutt (which typically have $M_B \sim
-21.3$ for Sab galaxies and $-20.9$ for Scs).  If we were to scale down
the star formation rate of a typical spiral by a factor of ten, we would have a
SFR of $\sim0.4$\,M$_\odot$/year.  Hence, scaling by optical luminosity, two
of our galaxies (ESO~118--G34 and NGC~2328) are forming stars at about the
same rate per unit blue luminosity as a typical spiral galaxy, while the
other two have a significantly lower SFR.

\subsection{Star formation history and the gas depletion time }

Most late--type disk galaxies have current star formation rates (SFR) which
are similar to the past rates averaged over the age of the disk (Kennicutt
1983).  In other word, these galaxies have evolved at a nearly constant rate.
Conversely, the star formation rates in luminous ellipticals were much higher
in the past than that they are now.  What can we say about the four galaxies
we have studied here?

To estimate the past star formation rate, we need to know both the total mass
and the mean age of the stellar population.  We have very little information
on the underlying stellar population other than the broad--band $UBV$ colours
listed in Table 1, but we can nevertheless set some limits on the {\sl
minimum} likely star formation rate in the past and get some indication of 
possible star formation histories. 

To make a rough estimate of the total {\sl stellar} mass for each galaxy, we
simply use the value of $L_B$ in Table 1 and assume $M/L_B = 3.0$.  
We first assume that all four galaxies began forming stars roughly
$10^{10}$ years ago (e.g.\ Searle et al.\ 1973).  Dividing the total stellar
mass by $10^{10}$ years implies a mean past SFR in the range $\langle R \rangle
= 0.6$-$4$ \msun/yr for the four galaxies.  In all four cases, this is
significantly higher than the current rates of 0.02 -- 0.3 \msun/yr.  The ratio
of present to past rates, SFR/$\langle R \rangle$ (Table 8) lies between 0.02
and 0.5.  Thus these galaxies appear to have had a different star--formation
history from the late--type disks studied by Kennicutt (1983). 

Searle et al.\ (1973) present model broad--band colours for galaxies which 
are $10^{10}$ years old and have an exponentially--declining SFR ($\propto
e^{-\beta t}$).  The broad--band colours in Table 1 are consistent with their
models for a Salpeter IMF and $\beta \sim 1$ if $t$ is in units of $10^{10}$
yr.  This is a very slow decline in SFR compared to the giant ellipticals,
whose colours are consistent with $\beta > 10$. 

The $UBV$ colours of the galaxies studied here, along with the current
star--formation rate estimated from H$\alpha$ emission lines, are consistent
with these low--luminosity star--forming E/S0 and S0
galaxies having a star formation history intermediate between 
late--type spiral disks (which have a roughly constant SFR) and giant
ellipticals (which have had little or no recent star formation). 
However, it is also important to point out that within our small sample we 
see considerable variation in the current SFR.  The two most `active' star 
formers, NGC~2328 and ESO~118--G34 have a SFR 5--15 times higher than 
NGC~802 and ESO~027--G21, even though all four galaxies have a similar 
\hi\ content. 

Column 7 of Table 8 gives the gas depletion time for the four galaxies, i.e.\ 
the time after which the total supply of \hi\ will be exhausted if star
formation continues at the present rate.  For the two galaxies with lower SFRs,
the gas depletion time is very long, whereas in NGC~2328 and ESO~118--G34 it is
of order 10$^9$ years (i.e.\ comparable to the value in many spiral disks;
Kennicutt 1983).  Note, however, that these estimates assume that there is 
no recycling of the ISM through stellar mass loss.  Kennicutt, Tamblyn \&
Congdon (1994) show that including a proper time--dependent treatment of the
gas return from stars extends the gas lifetimes of disks by factors of between
1.5 and 4.  This would imply gas depletion times of several Gyr even for 
NGC~2328 and ESO~118--G34.  

The discussion above assumes that the \hi\ disks in these galaxies are 
long--lived so that star formation can continue over a Hubble time.  
However, as discussed earlier, the \hi\ in two of our galaxies (NGC\,802 
and ESO\,118--G34) may have been accreted within the past 5$\times10^8$ years 
or so.  The current optical data do not rule out a small burst of recent star 
formation overlaid on a much older stellar population, and a more detailed 
study of the stellar population, measuring radial and line-strength and colour
gradients, would be useful to pin down the star formation history of these
galaxies more precisely.

\section{The FIR--radio correlation}

Since our four galaxies appear to be forming stars at a reasonable rate, 
it is interesting to see how they behave with respect to the well--known
correlation between the far-infrared (FIR) emission and the radio continuum
emission (the so-called FIR--radio correlation) observed for spiral
galaxies (e.g.\ Helou et al.\ 1985; Wunderlich et al.\ 1987, see Condon 1992
for a review).

The slope of the FIR--radio correlation is somewhat steeper than 1,
especially if low-luminosity galaxies are included in the samples studied
(Fitt et al.\ 1988, Cox et al.\ 1988, Devereux \& Eales 1989).
The properties of the four galaxies we studied are also consistent with a 
steeper slope. 
This can be quantified by  calculating the FIR/radio ratio $q$ 
at 1.4\,GHz following Helou et al.\ (1985): 
 $$
q = \log [ (S_{\rm FIR}/3.75\times 10^{12}\,{\rm Hz}) /S_{1.4} ] 
 $$ 
where $S_{\rm FIR} = 1.26\times10^{-14} [2.58 \cdot S_{60} + S_{100}]$, with
$S_{60}$ and $S_{100}$ the IRAS flux densities in Jy at 60 and 100\,$\mu$m
respectively and $S_{1.4}$ is the continuum flux density at 1.4\,GHz in W
m$^{-2}$ Hz$^{-1}$ (1 Jy = $10^{-26}$ W m$^{-2}$ Hz$^{-1}$).  The calculated
values of $q$ are given in Table 7.  The median value of $q$ for our galaxies
is $2.60$.  The values of $q$ are significantly higher than the mean value of
$2.14\pm 0.14$ measured for a sample of 38 more luminous spiral galaxies by
Helou et al.\ (1985). This means that they lie below a FIR--radio correlation
with slope 1, as used by Helou et al.

Our galaxies do follow the steeper correlation found by e.g.\ Devereux and
Eales (1989). As shown in Figure 9, all the galaxies (except NGC~1705, as
noted by Meurer et al.\ 1998) turn out to have a radio luminosity very close
to that predicted from their FIR luminosity (represented by the solid line).
Thus  the galaxies studied here are similar to spiral galaxies with a
similarly low FIR luminosity.  In principle it is possible that we could have 
underestimated the radio continuum flux densities by a factor of 2--3 by
resolving out very extended emission.  However, this would require most of the
radio continuum emission to be much more extended than the optical galaxy and
probably more extended than the \HI\ disk as well, which seems very unlikely. 

One explanation for the steeper slope of the FIR--radio correlation is that
the FIR emission consists of two components. One component is directly related
to the on-going star formation, while the other is a `cirrus'
component more related to the \HI\ emission of the disk (e.g.\ Devereux and
Eales 1989). In galaxies of low luminosity the contribution of the 
second component could be higher than in more luminous galaxies, so that 
low--luminosity galaxies would appear over--luminous in the FIR compared to
their radio flux. An alternative explanation is that the radio luminosity might
be deficient in low--luminosity galaxies because the cosmic rays are more
likely to escape by diffusion (Chi and Wolfendale 1990). 

In the four galaxies studied here, there is direct evidence for massive star
formation since \HII\ regions are seen.  We also know that the dust content of
these galaxies is relatively low (at least compared to their neutral gas
content -- see \S5 and Table 6) and the FIR dust temperature (35--40\,K; see
Table 7) similar to that seen in typical S0s and spirals.  Thus the most 
plausible explanation for the observed high values of $q$ appears to be 
a deficit of radio continuum emission rather than an FIR excess.  

The radio continuum 
emission from spiral galaxies has three main components (e.g.\ Condon 1992):\\ 
\noindent 
1.  Thermal (free--free) emission from \HII\ regions \\
2.  Non--thermal synchrotron emission from supernova remnants (SNRs) \\
3.  Large--scale non--thermal emission associated with the galaxy's disk. 

In spiral galaxies, radio emission from SNRs contributes no more than 3--10\%
of the total emission (e.g.\ Helou et al.\ 1985), and non--thermal disk
emission is always the dominant component (though its relative contribution
may vary with frequency).  Figure 10 plots H$\alpha$ luminosity 
$L_{{\rm H}\alpha}$
versus 1.4\,GHz radio power $P_{1.4}$.  For three of the four galaxies (the
exception is ESO\,027--G21), the entire radio continuum output can be
accounted for by thermal (free--free) radio emission from the HII regions
which produce the H$\alpha$ photons.  It therefore appears that these three
galaxies have no significant non--thermal disk emission in the radio
continuum, probably due to the absence of the magnetic field necessary to
generate synchrotron emission.

The remaining galaxy, ESO\,027--G21 (which is also the brightest of the four),
has radio continuum emission well in excess of that expected from the thermal
emission of its HII regions.  This galaxy nevertheless has a high $q$, so it
may be that non--thermal disk emission is present but at a lower level than
would be expected for a spiral galaxy.

\section{Discussion and Conclusions}

\subsection{\HI\ characteristics}

The observations presented in this paper confirm that if \HI\ is present in 
a low--luminosity early-type galaxy, it usually lies in a disk--like
structure. These \HI\ structures show, in varying degrees, regular rotation
while their density profiles are centrally concentrated. The central \HI\
densities are high enough for star formation to occur. Our observations bring
the number of low--luminosity early--type galaxies for which imaging \HI\ data
are available to about 10.  In almost all of these galaxies the \HI\ lies 
in disks or disk--like structures with high central densities. In
agreement with the findings of e.g.\ Lake et al.\ (1987), our data suggest
that in some galaxies a significant fraction of the \HI\ has been
accreted relatively recently. In other galaxies the distribution and
kinematics of the \HI\ do not show signs of any recent accretion event and the
gas must have been present in the galaxy, and in a disk, for a significant
time.

These characteristics differ from what is observed in more luminous early--type
galaxies. One main difference appears to be that central densities similar to
those observed in low--luminosity early--type galaxies (and the associated star
formation) are seldom found in more luminous early--type galaxies, even if they
have a regular \HI\ disk (e.g.\ Oosterloo et al.\ 1999a, 1999b). The peak 
\HI\ surface densities in luminous early--type galaxies are generally below 2
\msun ${\rm pc}^{-2}$ and the locations of the maximum surface densities are
usually at large radius and not near the center. In fact, the \HI\ distribution
in such galaxies often shows a central hole (e.g.\ Lake et al.\ 1987).  In
contrast, the central surface densities in low--luminosity early--type galaxies
appear to be more similar to those in late--type spirals.  The \HI\ surface
densities at larger radii do seem to be similar in low--luminosity and in
luminous early--type galaxies. 

Another difference appears to be that the {\sl range} in morphology and
kinematics of the \HI\ in such galaxies is larger than in low--luminosity
galaxies. More or less regular \HI\ structures have been observed in some
luminous early--type galaxies (e.g.\ NGC 807 Dressel 1987, NGC 2974 Kim et al.\
1988, NGC 3108 Morganti et al.\ 1999), but in many luminous early--type
galaxies the \HI\ has quite an irregular morphology, where the \HI\ usually
consists of filaments at large radii (e.g.\ NGC 2865 Schiminovich et al.\
1995; IC 1459 Oosterloo et al.\ 1999a). Counterparts of such cases do not
appear to exist at lower luminosity, or are at least much less common.

In general, the \HI\ content of luminous early--type galaxies depends on 
(at least) three factors (e.g.\ Knapp et al.\ 1985, Roberts et al.\ 1991, 
van Gorkom and Schiminovich 1997).  Since the \HI\ content correlates with the
presence of optical peculiarities, the 
\HI\ must have an external origin in many early-type galaxies and is the
remnant of an accretion/merger event. However, the \HI\ content is also an
indicator of how `disky' an early--type galaxy is because it also correlates
with the presence of a (usually faint) disk component. It is still conceivable
that in galaxies of this latter group the \HI\ also is accreted, but this then
must have happened a very long time ago allowing the gas to settle in a disk.
Alternatively, the \HI\ may have been accreted in a number of small accretion
events over a long period that would not disrupt the galaxy and would allow a
disk to form. A third factor is that the \HI\ content depends on environment.
 
Although, as in luminous early--type galaxies, evidence is observed that both
accretion and the presence of a disk play a role in low--luminosity early--type
galaxies, it appears that low--luminosity galaxies mostly belong to the 
second class of objects. This indicates that the evolution of these
systems is in general somewhat different than that of many luminous early--type
galaxies. The fact that in those low--luminosity early--type galaxies where the
\HI\ properties indicate that a significant fraction of the
\HI\ is accreted, the \HI\ structures are still  relatively
regular, suggests that the \HI\ has been accreted through less violent events
than in early--type galaxies of higher luminosity. 

The different \HI\ properties parallel the structural differences that are
observed at other wavelengths between low--luminosity and luminous early--type
galaxies, and that also point to a different evolutionary path for
low-luminosity galaxies.  For example, in low--luminosity early--type galaxies
the stellar rotation is in general more important than in more luminous 
early--type galaxies (e.g.\ Kormendy \& Bender 1996). It has recently been
suggested that this argues against major mergers as the dominant mechanism in
the final shaping of low--luminosity early--type galaxies galaxies and favors 
instead the dissipative formation of a disk (Rix, Carollo and Freeman, 1999).
The lack of chaotic \HI\ structures in low--luminosity early--type galaxies
would be consistent with this, as would perhaps the high central \HI\
densities. 

Studying a sample of merging galaxies where the progenitors are of roughly
equal mass, Hibbard and van Gorkom (1997) found that although large amounts of
\HI\ are often present in these systems, by the final stages there is little if
any \HI\ within the remnant body. They suggest that in such mergers the atomic
gas is efficiently converted into other forms within the main body of the
merger remnants. The \HI\ still present in these systems is likely to settle at
large radii.  The high \HI\ concentrations in low--luminosity early-type
galaxies would then suggest that no major merging event has occurred in these
galaxies. Alternatively (or additionally), it is also conceivable that the
centers of luminous early--type galaxies are too hostile for \HI\ to exist.

Optical emission lines of ionized gas have been observed in many luminous
early--type galaxies (e.g. Phillips et al.\ 1986).  The emission--line spectra 
are usually of the LINER class and are not \HII\ region--like spectra. This
indicates that gas is present, but that some mechanism prevents this gas from
being neutral. Mechanisms that have been suggested for this are photoionization
by old hot stars and mechanical energy from electron conduction by hot, X--ray
emitting gas (e.g.\ Goudfrooij 1999). It is possible, especially if hot 
X--ray gas plays a role, that this ionization mechanism does not work in
low--luminosity galaxies, so that central concentrations of \HI\ can still 
build up over time. 

\subsection{Star formation}

Consistent with the high central \HI\ surface densities in the galaxies studied 
here, star formation is occurring in their central regions.  A comparison of
past and present star formation rates suggests that these galaxies could have 
been forming stars steadily, at a slowly declining rate, for the past 10$^{10}$
years.  The question of whether these galaxies contain a component of
primordial \hi\ gas therefore remains open. A slowly declining star formation
rate would imply that these galaxies have contained significant amounts of
\HI\ during most of their existence, but in some galaxies there are also 
indications that \HI\ has recently been accreted.  A more detailed study of
the stellar population would be interesting if it could date the star
formation history more accurately.  The four galaxies show considerable
variation in their current SFR.  The two most `active' star formers, NGC~2328
and ESO~118--G34 have a SFR 5--15 times higher than NGC~802 and ESO~027--G21,
even though all four galaxies have a similar gas content.

Star formation appears to be limited to the central regions of the
galaxies.  Our observations are not detailed enough to allow the derivation of
the rotation curves of these galaxies. Hence, we cannot do an analysis of the
stability of the disks to see whether the star formation is occurring only
where it is expected to be. However, we can make an analogy with the starburst
dwarf galaxy NGC 1705, which shows many similarities with the galaxies
studied here (e.g.\ star formation only in the central regions).  Analysing
the stability of the disk of this galaxy, Meurer et al.\ (1998) found that 
star formation takes place where the disk is least stable. In NGC 1705
this is only near the center due to the central concentration of the
gas. Outside the central region, the surface densities are much lower and the
disk becomes stable. It is quite likely that in our four galaxies a similar
explanation applies.

Low--luminosity early--type galaxies appear to have similar ratios between the
FIR and radio fluxes as spirals of comparable FIR luminosity.  Many luminous
early--type galaxies deviate from this because their radio emission is
dominated by the contribution of an AGN (e.g.\ Wrobel \& Heeschen 1991, Bregman
et al.\ 1992). However such AGN are not present in the galaxies studied here,
which are expected to follow the FIR--radio correlation.  We suggest that the
galaxies observed here are underluminous in the radio continuum compared to
more luminous spirals, rather than overluminous in the FIR. The observed radio
continuum, except in ESO 027--G21, can be accounted for by thermal emission
alone. This suggests that the non--thermal emission, normally the strongest
component, is suppressed, and might indicate that cosmic rays are more likely
to escape the galaxy by diffusion, possibly due to a weaker magnetic field. 

\acknowledgements 

This research has made use of the NASA/IPAC Extragalactic Database
(NED) which is operated by the Jet Propulsion Laboratory, Caltech,
under contract with NASA.  We thank the anonymous referee for several 
helpful comments.

\clearpage

\figcaption[]{Single--dish \hi\ profiles measured at Parkes.}\label{parkes.ps}

\figcaption[] {Total \HI\ images (contours) obtained with natural
weighting superimposed to optical images from the Digital Sky Survey (DSS)
for NGC~802, ESO~118--G34, NGC~2328 and ESO~027--G21.  Contour levels for
NGC~802 from $2.9 \times 10^{19}$ \atms\ to $5.8 \times 10^{20}$ \atms\ in
steps of $4.3 \times 10^{19}$ \atms.  Contour levels for ESO~118-G34 from $3.1
\times 10^{19}$ \atms\ to $3.1 \times 10^{20}$ \atms\ in steps of $2.4 \times
10^{19}$ \atms.  Contour levels for NGC~2328 from $1.0 \times 10^{19}$ \atms\
to $2.0 \times 10^{20}$ \atms\ in steps of $2.0 \times 10^{19}$ \atms. 
Contour levels for ESO~027-G21 from $2.4 \times 10^{19}$ \atms\ to $8.1 \times
10^{20}$ \atms\ in steps of $8.1 \times 10^{19}$ \atms.  \\
The linear scale (for $H_\circ$ = 50 \kmsMp) is as follows: 
1\,arcmin = 7.7\,kpc for NGC\,802; 
1\,arcmin = 5.5\,kpc for ESO\,118--G34; 
1\,arcmin = 5.2\,kpc for NGC\,2328; 
1\,arcmin = 13.5\,kpc for ESO\,027--G21. 
\label{fig1}}

\figcaption[]{\HI\ velocity fields (contours) obtained with natural wieghting
superimposed to optical images from the Digital Sky Survey (DSS).  The
velocity contours are in steps of 5 \kms\ for NGC~802 ESO~118--G34 and
NGC~2328 and in steps of 10 \kms\ for ESO~027--G21.  \label{fig2}}

\figcaption[]{Total \HI\ and velocity fields (contours) obtained from the high
resolution data (see Table 4) superimposed to optical images from the Digital
Sky Survey (DSS) for NGC~802, ESO~118--G34, NGC~2328 and ESO~027--G21
(velocity contours as in Fig.3).  Contour levels for NGC~802 from $1.13 \times
10^{20}$ \atms\ to $1.13 \times 10^{21}$ \atms\ in steps of $1.13 \times
10^{20}$ \atms.  Contour levels for ESO~118-G34 from $6.1 \times 10^{19}$
\atms\ to $6.1 \times 10^{20}$ \atms\ in steps of $6.1 \times 10^{19}$ \atms. 
Contour levels for NGC~2328 from $2.5 \times 10^{19}$ \atms\ to $2.5 \times
10^{20}$ \atms\ in steps of $2.5 \times 10^{19}$ \atms.  Contour levels for
ESO~027-G21 from $6.0 \times 10^{19}$ \atms\ to $6.0 \times 10^{20}$ \atms\ in
steps of $6.0 \times 10^{19}$ \atms.  \label{fig3}}

\figcaption[]{Position-velocity plots for ({\sl top})
NGC~802 low and high resolution in p.a. 90$^\circ$; ({\sl middle})
ESO~118--G34 in p.a. 160$^\circ$ and NGC~2328 in p.a. 110$^\circ$; 
({\sl low)} ESO~027--G21 in p.a. 80$^\circ$.
Contour levels for NGC~802 low resolution from 2 to 23 \mJybeam\ in step of
2\mJybeam and 
from 2.5 to 13 \mJybeam\ in step of 1.5\mJybeam\ for the high resolution. 
Contour levels for ESO~118-G34 from 2.5 to 15 \mJybeam\ in step of 1\mJybeam.
Contour levels for NGC~2328 from 2.0 to 11 \mJybeam\ in step of 1\mJybeam.
Contour levels for ESO~027-G21 from 3.0 to 30 \mJybeam\ in step of 2.5\mJybeam.
\label{fig5}}

\figcaption[]{Optical spectra and rotation curves for {\sl (a)} NGC~802 and 
{\sl (b)} ESO~027--G21.\label{fig6}}

\figcaption[]{Radial \HI\ surface density profiles for the four galaxies.
For all four galaxies, crosses show the profiles measured from 
natural--weighting maps.  For NGC~802 and ESO~118--G34, the open circles show
the profiles derived from higher--resolution observations.  For NGC\,2328 and
ESO\,027--G21, open circles show profiles measured from maps made with natural
rather than uniform weighting. The vertical dashed lines show the outer limit
of star--forming regions in each galaxy.}\label{surf.ps} 

\figcaption[]{Total \HI\ of images for two companion galaxies:
{\sl Top:} AM~0155--675 in the field of NGC~802. Contour levels from $1.13
\times 10^{20}$ \atms\ to $4.52 \times 10^{20}$ \atms\ in steps of $1.13
\times 10^{20}$ \atms.  {\sl Bottom:} anonymous galaxy in the field of
ESO~027--G21. Contour levels from $2.2 \times 10^{19}$ \atms\ to $8.0 \times
10^{20}$ \atms\ in steps of $6.0 \times 10^{19}$ \atms. }

\figcaption[]{The FIR--radio correlation for four galaxies observed here, 
together with the four small \hi--rich ellipticals observed by Lake et al.\ 
(1987) (NGC~3265, NGC~5666, UGC~7354 and A1230$+$09) and the blue compact 
galaxy NGC~1705 (Meurer et al.\ 1998).  The solid line shows the relation 
derived for  spirals by Devereux and Eales (1989), which has a slope of 
1.28. \label{fig9}}

\figcaption[]{Comparison of H$_\alpha$ emission--line luminosity and
radio continuum flux density for the four galaxies (the radio continuum point
for NGC~802 is an upper limit).  The solid line shows the relation expected for
\HII\ regions (Terzian 1965). \label{fig10}}


\begin{table}
\caption{Optical properties of the four galaxies } 
\begin{tabular}{lccccc}
\tableline\tableline
&{\bf  NGC\,802} & {\bf ESO\,118--G34}& {\bf NGC\,2328}& {\bf ESO\,027--G21} &\\
\tableline
RA (J2000.0)   & 01 59 07.1    &    04 40 17.2  &  07 02 36.7 &23 04 19.6 & a \\
Dec            &  $-$67 52 11  & $-$58 44 47    &$-$42 04 09  & $-$79 28 01& \\
Type           &  E/S0         &    S0          &    E/S0      &    S0    & b  \\
$B_{\rm T}^{\ 0}$&  14.12      &   13.49        &    12.54    &  13.38   & a \\
$(B-V)_{\rm T}^{\ 0}$ & 0.43  &    0.45        &   0.49      &   --    & a \\
$(U-B)_{\rm T}^{\ 0}$ &  $-$0.15 & $-$0.36     &   0.09      &   --    & a \\
$r_{\rm e}$ (arcsec,kpc) & 16.5(2.1)  &  13.1(1.2)     & 17.3(1.5)   & 19.0(4.3)  & c \\
$v_{\rm hel}$ (\kms)   &  1505         &   1171         &    1159     &  2504    & a \\
$v_{\rm hel}$ (\kms)   &  $1475\pm 5$ &   $1150\pm 5$  &  $1168\pm 5$ & $2480\pm 5$  & d \\
$v_\circ$ (\kms)   &  1321         &    941         &     884     &  2316    &   \\
Distance (Mpc) &  26.4         &   18.8         &    17.7     &  46.3    & \\
$M_{B}$ (mag)  &  $-$18.0      &   $-$17.9      &   $-$18.7   &  $-$19.9 & \\
$L_{B}$ (\lsun) &  $2.2\times 10^9$ & $2.1\times 10^9$& $4.4\times 10^9$&  
 $1.4\times 10^{10}$ & \\
\tableline\tableline
\end{tabular}
\tablecomments{References: a) revised RC3  b) ESO/Uppsala cat. c) Sadler (1984) d) this paper}
\end{table}

\clearpage

\begin{table}
\caption{Single--dish \HI\ observations at Parkes}
\begin{tabular}{lcccccc} \tableline\tableline
{Galaxy}&   {V$_{\rm central}$} &  Range & {On--source}& Total \HI\ flux 
& FWHM  & \HI\ mass \\ 
         &      (\kms)             & (\kms) & {time} (min)& (Jy \kms)  
& (\kms) & (M$_\odot$)   \\
 \hline
 NGC\,802      & 1475 &  500--2650 & 20 & 5.9 &  73 & 9.7$\times10^8$ \\
 NGC\,2328     & 1167 &  695--1760 & 30 & 4.4 & 111 & 3.3$\times10^8$ \\ 
 ESO\,027-G21  & 2480 & 1530--3650 & 40 &10.8 & 154 & 5.5$\times10^9$ \\
\tableline\tableline
\end{tabular}
\end{table}

\clearpage

\begin{table}
\caption{Log of ATCA Observations}
\begin{tabular}{cccclc} \tableline \tableline
{Galaxy} &{Date} &   {Configuration} & {Baseline} &  {Frequency} & {Time} \\ 
         &       &                   & {range} (m)&   (MHz)      &     \\
 \hline
 NGC\,802 & Jan95   & 375   &  31--459      & 1413 (1380) &  12h \\
 NGC\,802 & Feb96   & 1.5C  &  77--1485     & 1413 (1346) &  12h \\
 ESO\,118--G34 & Dec96 & 375 & 31--337      & 1415 (1344) &  12h \\
 ESO\,118--G34 & Mar97 & 1.5D& 214--1439    & 1415 (2396) &  12h \\
 NGC\,2328 & Nov96  & 750A  &  77--735      & 1415 (1344) & 12h \\ 
 ESO\,027-G21 & Nov96 & 750A &  77--735     & 1409 (1344) & 12h \\
\tableline\tableline
\end{tabular}
\end{table}

\clearpage

\begin{table}
\caption{Instrumental Parameters of the ATCA observations}
\begin{tabular}{lcccc} \tableline\tableline
{\bf \ion{H}{1} observations}  &&&&  \\ 
 & {\bf NGC\,802} & {\bf ESO\,118--G24} &{\bf NGC\,2328} &{\bf ESO\,027--G21}\\
\tableline
RA (J2000.0)   & 01 59 07    &    04 40 08  &  07 02 36  & 23 04 19  \\
Dec            &  $-$67 52 40  & $-$58 45 59    &$-$42 04 09  & $-$79 30 01 \\
Synthesized beam &
$68.1^{\prime\prime} \times 54.6^{\prime\prime}$ & 
$60.9^{\prime\prime} \times 58.0^{\prime\prime}$ & 
$79.7^{\prime\prime} \times 56.4^{\prime\prime}$ & 
$64.8^{\prime\prime} \times 49.2^{\prime\prime}$ \\
(natural weighting) & & & & \\
Beam p.a.   & $-11^\circ$ & $0^\circ$ & $6^\circ$ & $-45^\circ$ \\
Synthesized beam &
$37.0^{\prime\prime} \times 25.0^{\prime\prime}$ & 
$25.0^{\prime\prime} \times 23.0^{\prime\prime}$ & 
$56.0^{\prime\prime} \times 41.0^{\prime\prime}$ & 
$54.3^{\prime\prime} \times 40.4^{\prime\prime}$  \\
(weighting) & robust=0 & uniform  & uniform  & robust=0  \\
Beam p.a.   & $0^\circ$ & $-2^\circ$ & $0^\circ$ & $-40^\circ$ \\
Number of channels          & 512   & 512 & 512 & 512  \\
Central velocity (\kms)     & 1580  & 1150 & 1150 & 2420   \\
Velocity resolution         &  6.0  & 6.0  &   6.6  &  6.6     \\
(after Hanning) (\kms)            &       &      &      &        \\
rms noise in channel maps    &  1.1  &  1.5   & 1.2 & 1.5 \\
(nat.\ weight, \mJybeam)   &       &        &     &     \\
\tableline\tableline
\end{tabular}
\end{table}

\clearpage

\begin{table}
\caption{Radio continuum observations of the four galaxies.}
\begin{tabular}{lcccrlcl} \tableline\tableline
Galaxy & Array & Freq. & Beam, PA & \multicolumn{2}{c}{Flux density (mJy)}& rms 
& Notes \\
       &        & (GHz)& (arcsec) &   Peak &  Total & (mJy) &   \\
NGC\,802 & 375$+$1.5 & 1.3  & $24.1\times20.6$ (24) & $<$1.9 & --   & 0.37 & 
          Not detected \\
         &  1.5C& 1.3  &$7.5\times5.4$ (-75)& $<$1.0  & --  & 0.19 & 
          6~km ant. \\
ESO\,118--G34 & 375$+$1.5& 1.3& $20.3\times18.5$ (-2) & 4.0 & 7.0 & 0.21 & \\
              &1.5D & 1.3& $8.9\times8.3$ (-7)   & 2.8 & 6.5: & 0.36 &  
           6~km ant. \\ 
NGC\,2328 & 750A & 1.3 & $55.9\times41.6$ (6) & 8.1 & 9.4 & 0.13 & \\
          & 750A & 1.3 & 8.3$\times$6.2 (5)  & 3.1 & 8.3 & 0.11 & 6~km ant. \\
ESO\,027--G21 & 750A & 1.3 & 42.0$\times$39.8 (42) & 4.8 & 4.7 & 0.20 & \\
          & 750A & 1.3 & 6.2$\times$5.9 (-41)  & 1.5 & 3.7 & 0.10 & 6~km ant. \\
\hline
ESO\,118--G34 & 375$+$1.5 & 2.5& $17.5\times14.9$ (-8) & 3.1 & 4.8 & 0.05 & \\
           & 1.5D & 2.4 & $10.1\times8.8$ (-1) & 1.9 & 4.4 & 0.07& 6~km ant. \\
\tableline\tableline
\end{tabular}
\end{table}

\clearpage
 
\begin{table}
\caption{Components of the ISM in the four galaxies.}
\begin{tabular}{lccccc} \tableline\tableline
&{\bf  NGC\,802} & {\bf ESO\,118--G34}& {\bf NGC\,2328}& {\bf ESO\,027--G21} &\\
$M_{\rm HI}$ (\msun)  & $9.7\times 10^8$ &  -- & $3.3\times 10^8$& 
$5.5\times 10^9$ &  a\\
$M_{\rm HI}$ (\msun)  & $7.6\times 10^8$ & $2.5\times 10^8$& $2.0\times 10^8$ & 
 $4.7\times 10^9$ &   b\\
$M_{\rm HI}/L_B$      &  0.42 & 0.12 & 0.07  &  0.40 &   \\
\hline
$M_{\rm H2}$ (\msun)  &  &   & $2.9\times 10^8$ &  &  c \\
$M_{\rm H2}$/M$_{\rm HI}$ & &  & 0.87 &  & \\
\hline
$L_{{\rm H}\alpha}$ (W) &  $2.3\times 10^{32}$ & $3.2\times 10^{33}$ & 
 $3.6\times 10^{33}$&  $7.8\times 10^{32}$ & \\
$M_{\rm HII}$ (\msun)  & $5.5\times 10^3$ &  $7.3\times 10^4$& 
 $8.3\times 10^4$ &  $1.8\times 10^4$ &   \\
\hline
$S_{60\mu{\rm m}}$ (mJy)   & 380 & 2090       &  2770    &   780 & d \\
$S_{100\mu{\rm m}}$ (mJy)  & 830 & 3190       &  3770    &  1800 & d \\
$L_{{\rm FIR}}$ (\lsun)    & $4.6\times 10^8$ & $1.1\times 10^9$ & 
 $1.3\times 10^9$ & $3.0\times 10^9$ &   \\
$M_{\rm dust}$ (\msun)  & $1.5\times 10^5$ & $1.7\times 10^5$ & 
 $1.5\times 10^5$ & $1.1\times 10^6$ &   \\
$M_{\rm HI}/M_{\rm dust}$ & 6415 & 1456 & 2144 & 4944 & \\
\tableline\tableline
\end{tabular}
\tablecomments{(a) From single--dish data, (b) from ATCA data (c) Wiklind 
et al.\ (1995) (d) Knapp et al. (1989).}  
\end{table}

\clearpage

\begin{table}
\caption{Star formation properties of the four galaxies.}
\begin{tabular}{lcccc} \hline \hline
      & {\bf NGC\,802} & {\bf ESO\,118--G34} & {\bf NGC\,2328} & 
 {\bf ESO\,027--G21}  \\
SFR ($>10$\,\msun) (\msun/yr)  & 0.003  & 0.046 & 0.051 & 0.011 \\
SFR (total) (\msun/yr)         & 0.021  & 0.29  & 0.32  & 0.070 \\
SFR (rel. to NGC\,802)         & 1.0    & 13.8  & 15.2  & 3.3   \\
\hline 
$\log P_{1.4}$(total) (W/Hz)      & $<$19.93 & 20.47 & 20.55 & 21.09 \\
FIR/radio ratio $q$            & $>$2.78 & 2.61 & 2.59  & 2.43  \\
$L_{\rm FIR}/L_B$         & 0.20 & 0.53 & 0.28 & 0.22         \\
$T_{60/100\mu \rm m}$     & 36\,K & 41\,K & 43\,K & 35\,K \\
\tableline\tableline
\end{tabular}
\end{table}

\clearpage

\begin{table}
\caption{Timescales for star formation.}
\begin{tabular}{lcccccclc} \tableline\tableline
 Galaxy   &  SFR & $T_{\rm dust}$ & $M_{\rm HI}/L_B$ & Est.\ $M_{\rm stars}$ &  SFR & $T$(\HI\ depletion)  \\
          & ($M_\odot$/yr) & (K)  &    & ($M_\odot$) & /$\langle R \rangle$ 
&  (yr) \\ 
NGC\,802  &  0.021  & 36 &  0.42  & 6.6$\times10^{9}$ & 0.032 & $7\times 10^{10}$  \\ 
ESO~027--G21  &  0.070  & 35 &  0.40  & 4.2$\times10^{10}$ & 0.017 & 8$\times
10^{10}$  \\
\hline
ESO~118--G34  &  0.29   & 41 &  0.12  & 6.3$\times10^9$ & 0.46 & 8$\times 10^8$ 
\\ 
NGC\,2328 &  0.32   & 43 &  0.07  & 1.3$\times10^{10}$ & 0.25 & 2$\times 10^9$ 
  \\ 
\tableline\tableline
\end{tabular}
\end{table}

\end{document}